\documentclass[epsfig,useAMS,usenatbib]{mn2e}
\usepackage{epsfig,graphics}
\def\bib{\parskip=0pt\par\noindent\hangindent\parindent \parskip =2ex
  plus .5ex
 minus .1ex} \newcommand{\be}{\begin{equation}}
\newcommand{\ee}{\end{equation}} \title[The 2TecX algorithm]{A
  reliable cluster detection technique using photometric redshifts:
  introducing the 2TecX algorithm} \author[Van Breukelen \&
  Clewley]{Caroline van Breukelen$^{1,2}$\thanks{cvb@star.ucl.ac.uk}
  \& Lee Clewley$^{2}$ \\ $^{1}$Physics \& Astronomy, University
  College London, Gower Street, London, WC1E 6BT,
  UK\\ $^{2}$Astrophysics, Department of Physics, Keble Road, Oxford,
  OX1 3RH, UK}

\date{Released 2002 Xxxxx XX}

\pagerange{\pageref{firstpage}--\pageref{lastpage}} \pubyear{2002}

\def\LaTeX{L\kern-.36em\raise.3ex\hbox{a}\kern-.15em
    T\kern-.1667em\lower.7ex\hbox{E}\kern-.125emX}

\def\am{^{\prime}}
\def\deg{^{\circ}}
\def\grtsim{\mathrel{\hbox{\rlap{\hbox{\lower2pt\hbox{$\sim$}}}\raise2pt\hbox{$>$}}}}
\def\lesssim{\mathrel{\hbox{\rlap{\hbox{\lower2pt\hbox{$\sim$}}}\raise2pt\hbox{$<$}}}}

\begin{document}
\label{firstpage}
\maketitle
\begin{abstract}
We present a new cluster detection algorithm designed for finding
high-redshift clusters using optical/infrared imaging data. The
algorithm has two main characteristics. First, it utilises each
galaxy's full redshift probability function, instead of an estimate of
the photometric redshift based on the peak of the probability function
and an associated Gaussian error. Second, it identifies cluster
candidates through cross-checking the results of two substantially
different selection techniques (the name 2TecX representing the
cross-check of the two techniques). These are adaptations of the
Voronoi Tesselations and Friends-Of-Friends methods. Monte-Carlo
simulations of mock catalogues show that cross-checking the
cluster candidates found by the two techniques significantly reduces
the detection of spurious sources. Furthermore, we examine the
selection effects and relative strengths and weaknesses of either
method. The simulations also allow us to fine-tune the algorithm's
parameters, and define completeness and mass limit as a function of
redshift. We demonstrate that the algorithm isolates high-redshift
clusters at a high level of efficiency and low contamination.
\end{abstract}

\begin{keywords}
methods: data analysis -- galaxies: clusters: general -- techniques: photometric
 \end{keywords}

\section{Introduction}
Remote galaxy clusters have been used in a wide range of cosmological
and astrophysical contexts. In cosmology, clusters can be used to
trace the large-scale structure of the universe. Their number density,
as a function of redshift, can place constraints on various
cosmological quantities. These include the mass density of the universe, the
amplitude of the initial density fluctuations, and the cosmic growth
function. Clusters also act as astrophysical laboratories for
understanding the formation and evolution of galaxies and their
environments. This is because the deep potential well of a cluster
causes it to retain virtually all its gas and galaxies, allowing a
detailed inspection of the interaction between both. It is therefore
desirable to have a large, homogeneous catalogue of clusters at a
range of redshifts in the universe.

Abell compiled the first large cluster catalogue, in which clusters
were selected in a consistent manner (Abell 1958; Abell, Corwing \&
Olowin 1989). This catalogue was created from photographic
observations, which suffer from non-linear plate-to-plate sensitivity
variations and considerably large photometric errors (Sutherland
1988). Furthermore, the clusters were found by eye which poses
problems for the objectivity and completeness of the cluster sample
and the line-of-sight projections contaminating it (e.g. Lucey 1983;
van Haarlem, Frenk \& White 1997).  A particularly important advance
has come from optical galaxy surveys using large arrays of CCD
detectors, such as the relatively shallow (z $<$ 0.4) Sloan Digital
Sky Survey (SDSS) (e.g Goto et al. 2002; Kim et al. 2002; Miller et
al. 2005). A recent large-scale cluster catalogue using the SDSS was
initiated by Koester et al. (2007a,b), detecting $\sim 1400$ clusters
at $0.1 < z < 0.3$. There have been numerous smaller-area surveys to
much higher redshift, as for instance the Palomar Distant Cluster
Survey (Postman et al. 1996); the ESO Imaging Survey (Lobo et
al. 2000); and the Red Sequence Cluster Survey (Gladders \& Yee 2005).

Optical cluster surveys were limited for a long time to clusters at $z
\lesssim 1$, due to the fact that the cluster galaxy population largely
consists of early-type red galaxies. At redshifts of $z \grtsim 1$,
the 4000~\AA\ break moves into infrared bands, complicating the
detection of these galaxies in optical surveys. A crucial development
has been the advent of wide-field infrared cameras. Deep, large-area
infrared studies have already become available from the Wide Field
Infrared Camera (WFCAM) on the United Kingdom Infra-Red Telescope
(UKIRT) and the {\it Spitzer} space telescope and will shortly be available
on the Visible and Infrared Survey Telescope for Astronomy (VISTA).

There exist many methods for detecting clusters in optical imaging
surveys. The problem is somewhat easier for galaxy datasets with
spectroscopic redshifts owing to the accurate knowledge of each galaxies
distance. However, spectroscopy is time consuming and approximate
redshifts can be calculated via photometric redshift estimation. This
technique is considerably less precise which makes looking for
structure less straightforward. A successful photometric method for
finding clusters is to use deep optical imaging data that span the
rest frame 4000~\AA\ break (Gladders \& Yee, 2000). This is motivated
by the observation that cluster early-type galaxies form a
characteristic red sequence comprising the brightest, reddest galaxies
at a given redshift. The colour of this red sequence also provides an
estimate of the redshift of the detected cluster, thereby reducing
projection effects (e.g. Gladders and Yee, 2005). However, at high
redshift there is not yet substantial evidence whether all clusters do
indeed show a red sequence. Merely selecting by this characteristic
could be introducing a large bias against younger clusters with
ongoing star-formation.

In this paper we present a new cluster detection method, specifically
designed to detect high-redshift clusters using optical/infrared
imaging data. In Section 2 we describe the cluster detection algorithm
step by step. Section 3 contains details of the creation of mock
catalogues, along with simulations for parameter optimisation and to
determine the completeness and contamination by spurious
sources. Section 4 is a summary of the algorithm and its performance
on the set of simulations. We assume throughout this paper that $h =
H_0 / 100~{\rm km~s^{-1}Mpc^{-1}} = 0.7$, and a $\Omega_ {\rm M}=0.3$,
$\Omega_ {\Lambda}=0.7$ cosmology. All magnitudes are given in the
Vega system.

\section{2TecX: a new cluster detection algorithm} 

Optical cluster surveys using selection methods based on photometric
redshifts often suffer from two common problems: (i) projection
effects of fore- and back-ground galaxies and (ii) determining the
reality of detected clusters. The former issue arises because
photometric redshifts, as opposed to spectroscopic redshifts,
typically have errors of the order of $\sigma \sim 0.1$; furthermore
the photometric redshift probability functions (z-PDFs) are often
significantly non-Gaussian and can for instance show double peaks. The
second issue -- the occurrence of spurious cluster detections -- is
due to sensitivity of the detection algorithm to noisy data. To create
a cluster catalogue a compromise needs to be made between completeness
and contamination: we want to include as many clusters as possible
above a certain mass limit, without suffering from contamination by
spurious sources. It is important to understand the completeness and
efficiency of cluster finders.

To address these two problems, we create a new cluster-detection
algorithm that is characterised by two main improvements upon previous
work: (i) the cluster-detection algorithm utilises the full z-PDF
instead of a single best redshift-estimate with an associated Gaussian
error; (ii) we maximise the efficiency by cross-checking the output
of two substantially different cluster detection methods.

\begin{figure}
\begin{center}
\includegraphics[width=8cm]{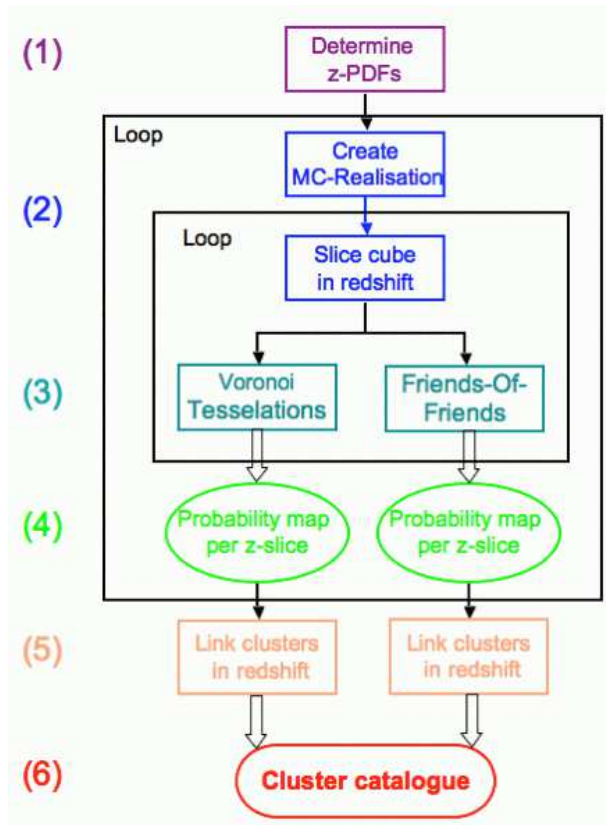}
\end{center}
\caption{\small Schematic diagram of the cluster-selection
  algorithm. Each of the steps is described in detail in
  Sections~\ref{z-pdf} to \ref{prob_maps}.}
\label{schema}
\end{figure}

The algorithm is divided into six steps, described in more detail in
the following subsections and shown schematically in
Fig.~\ref{schema}:
\begin{enumerate}
\item Determining z-PDFs for all galaxies in the field.
\item Creating 500 Monte-Carlo (MC) realisations of the three-dimensional
  galaxy distribution, based on the galaxy z-PDFs.
\item Dividing each MC-realisation into redshift slices of $\Delta z =
  0.05$ over the range $0.1 \leq z \leq 2.0$.
\item Detecting cluster candidates in each slice of all
  MC-realisations using independent Voronoi Tessellation (VT) and
  Friends-Of-Friends (FOF) methods.
\item Mapping the probability of cluster candidates for both methods
  based on the number of MC-realisations in which they occur.
\item Cross-checking the output of the VT and FOF methods to arrive
  at the final cluster-catalogue.
\end{enumerate}

\subsection{Redshift probability distribution functions}\label{z-pdf}

The photometric redshifts of Van Breukelen et al. (2006, henceforth
VB06), who first applied our cluster-detection algorithm to
optical/infrared imaging data, were created by an adapted version of
{\it Hyperz} (Bolzonella et al. 2000), using a set of Spectral Energy
Distributions (SEDs) generated with {\it GALAXEV} (Bruzual \& Charlot
2003). {\it Hyperz} estimates photometric redshifts by fitting a range
of SED templates to the measured fluxes in several photometric
bands. The shape of the SEDs are determined by various parameters,
such as the rate of ongoing star-formation, the age of the galaxy, the
metallicity, and the reddening due to extinction. A redshift
probability distribution function is constructed by calculating the
probability of the best-fitting set of parameters at each
redshift. Thus the z-PDF does not reflect the probability with
redshift for a single template, but rather for the total set of
templates. The location of the maximum of the z-PDF is taken as the
photometric redshift and an error can be estimated by fitting a
Gaussian profile to the probability peak. However, this does not take
into account the often non-Gaussian and sometimes double-peaked nature
of the z-PDF. These can arise because different features of the
spectrum can be confused (for example the 4000\,\AA\ break and the
Lyman-$\alpha$ break at $\sim 1000$\,\AA) or various templates can
give solutions of comparable probability at different redshifts. We
therefore do not use a best-estimate photometric redshift, but take
the entire z-PDF into account in our cluster search. The output of our
adapted {\it Hyperz} program is the marginalised likelihood associated
with each step in redshift space for each galaxy. However, the 2TecX
algorithm can be applied to any photometric redshift dataset that
contains a z-PDF for every galaxy. 

\subsection{The Monte-Carlo realisations and redshift slicing}\label{mc}
To include the entire z-PDF of each galaxy into our cluster-detection
algorithm, we create 500 MC-realisations of the three-dimensional
galaxy distribution by randomly sampling each z-PDF. We chose the
number of realisations as a compromise between computational time and
sampling accuracy of the z-PDF. We now have 500 cubes of RA,
Dec, and $z$, where each galaxy is represented by a single point. The
shape of the z-PDF of each galaxy determines its position in the
cubes; if the peak in the probability distribution function is sharp
the galaxy will occur in all cubes at approximately the same redshift
whereas if the z-PDF consists of two equally probable peaks the galaxy
will be placed at either redshift in an equal number of cubes.

Next, we divide each MC-realisation into redshift slices of a width,
$\Delta z$, approximately equal to the photometric redshift error,
$\sigma_z$. If the width is chosen to be significantly smaller,
clusters can be undetected due to the distribution of their member
galaxies over too many redshift slices; if it is chosen substantially
larger, many spurious sources will be found owing to projection
effects. In this paper we use $\Delta z = 0.05$, as this is the
approximate photometric redshift error of VB06.

\subsection{Two cluster selection methods}\label{methods2}
We now have 500 MC-realisations of the three-dimensional galaxy
distribution, each divided into redshift slices.  In the next step,
the algorithm applies two cluster selection methods independently to
each redshift slice of all the MC-realisations. The two methods used
are Voronoi Tessellation and Friends-Of-Friends, which are described
in more detail below.

\begin{figure}
\hspace{-0.5cm}
\includegraphics[width=9cm]{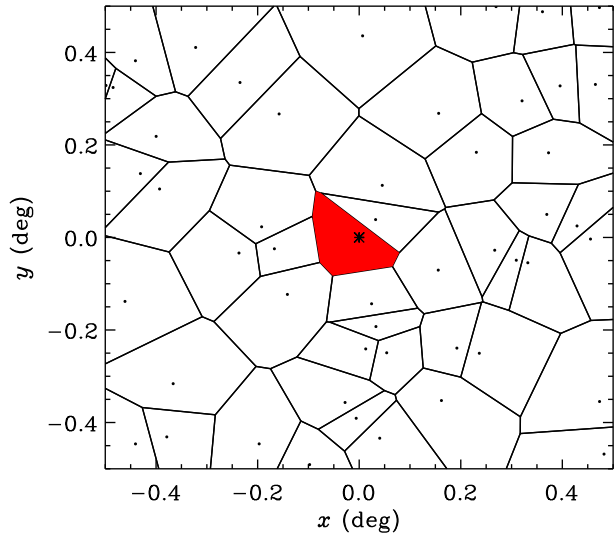}
\vspace{-0.5cm}
\caption{\small An example of Voronoi Tessellations. The dots
  represent the nuclei, randomly distributed over the field. Each
  Voronoi Cell encloses all points in the field that are closer to its
  nucleus than to any other nucleus. For example, all points within
  the filled (red) Voronoi Cell are closer to the nucleus marked by
  the star symbol than to any of the dots.}
\label{voronoi}
\end{figure}

\begin{figure}
\hspace{-0.5cm}
\includegraphics[width=9cm]{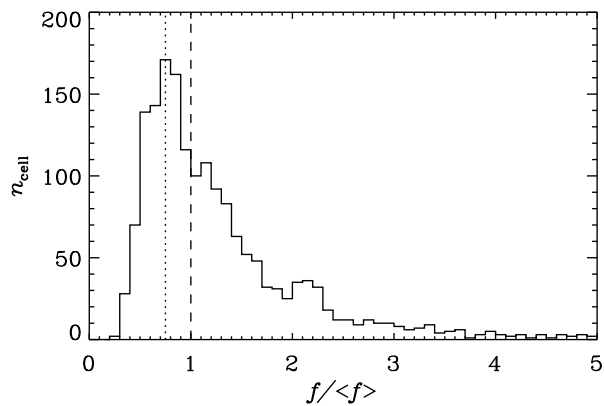}
\vspace{-0.5cm}
\caption{\small Histogram of Voronoi Cell densities in a field of 2000
  randomly distributed background galaxies including a central
  overdensity of 100 galaxies with a Gaussian density profile with
  $\sigma = 1\am$. The dashed line is placed at $f = <\!\!f\!\!>$ and
  the dotted line denotes the position of the peak which is at $f_{\rm
    max} = \frac{3}{4}<\!\!f\!\!>$.}
\label{vorhist}
\end{figure}

\subsubsection{Voronoi Tessellations}\label{VT}

\begin{figure*}
\begin{center}
\includegraphics[height=15cm,angle=90]{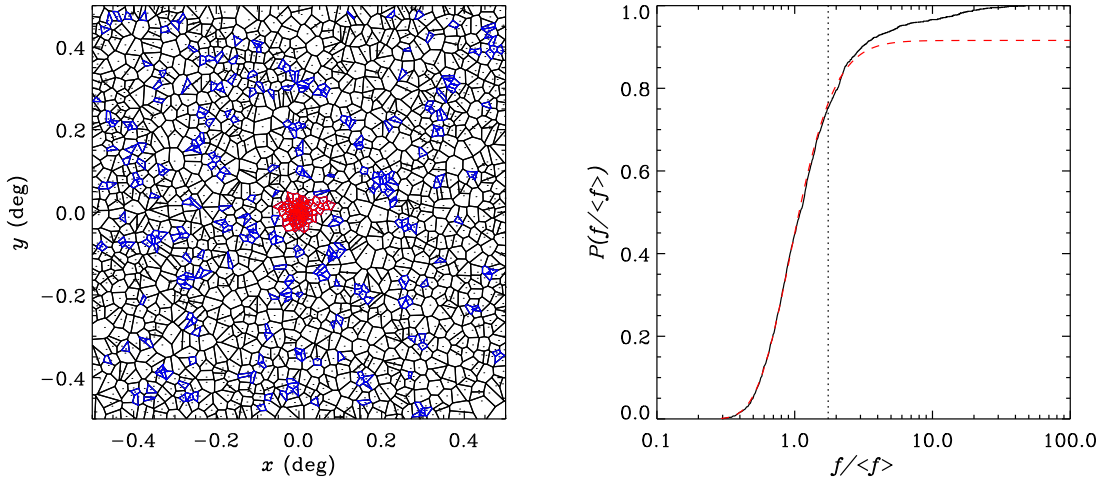}
\end{center}
\vspace{-0.5cm}
\caption{\small {\it Left:} Voronoi Tessellations on a field of
  background sources with a central overdensity superimposed. The
  background consists of 2000 galaxies uniformly distributed
  throughout the field. The central structure comprises 100 galaxies
  and has a Gaussian density distribution with a $\sigma = 1 \am$. The
  blue cells denote the cells with density $\tilde f > \tilde f_{\rm
    min}$. The red cells compose the group that also satisfies the
  $n_{\rm gal} > n_{\rm lim}$ criterion.  Note that all the
  high-density background fluctuations (blue cells) are not selected
  as cluster candidates. {\it Right:} The cumulative density
  distribution of the data in the field shown on the left. The red
  dashed line is the fit to the lower-density cells according to
  Eq.~\ref{cf_f}. The dotted vertical line shows the value of $\tilde
  f_{\rm min} = f/ <\!\!f\!\!>$, the minimum density above which
  high-density cells are selected (see Section~\ref{results2} for a
  discussion of the value of this parameter).}
\label{vor_func}
\end{figure*}

The VT technique divides a field of galaxies into Voronoi Cells, each
containing one object: the nucleus. All points that are closer to this
nucleus than any of the other nuclei are enclosed by the Voronoi Cell
(see Fig.~\ref{voronoi}). This technique was first applied to the
modelling of large-scale structure (e.g. Icke \& van de Weygaert 1987)
but has more recently been used in cluster detection (Ebeling \&
Wiedenmann 1993; Kim et al. 2002; Lopes et al. 2004). One of the
principal advantages of the VT method is that the technique is
relatively unbiased as it does not look for a particular source
geometry (e.g. Ramella 2001). The parameter of interest is the area of
the VT cells, the reciprocal of which translates to a
density. Overdense regions in the plane are found by fitting a
function to the density distribution of all VT cells in the field;
cluster candidates are the groups of cells of a significantly higher
density than the mean background density.

Kiang (1966) showed that, for randomly (Poissonian) distributed
points, the differential distribution function of the cell area is of
the following form: 
\be 
dp(\tilde a) = \frac{4^4}{\Gamma(4)}\tilde a^3
e^{-4\tilde a}d\tilde a.
\label{kiang}
\ee Here $\tilde a \equiv a~ / <\!\!a\!\!>$ is the dimensionless cell
area in units of the average cell area: $<\!\!a\!\!> =
\frac{1}{N}\sum^N_{i=1}a_i$, where $N$ is the total number of
cells. $\Gamma(x)$ is the Gamma Function. The cumulative distribution
function for the cell area $\tilde a$ is the integral of
Eq.~\ref{kiang}, namely: \be P(\tilde a) = 1 - e^{-4\tilde
  a}\Bigl(\frac{32\tilde a^3}{3} + 8\tilde a^2 + 4 \tilde a + 1\Bigr).
\label{cf_a}
\ee
The density of the VT cells is the reciprocal of Eq.~\ref{cf_a}:
\begin{equation}
P(\tilde f) = e^{-4/\tilde f} \Bigl (\frac{32}{3\tilde f^3} +
\frac{8}{\tilde f^2}+ \frac{4}{\tilde f} + 1 \Bigr ).
\label{cf_f}
\end{equation}
Here $\tilde f$ is the dimensionless cell density (the inverse of the cell area) in
units of the mean cell density: 
\be
\tilde f = f/<\!\!f\!\!> = <\!\!a\!\!>/a.
\label{tilde_f}
\ee In our algorithm, we approximate the density distribution of the
background galaxies by a Poissonian distribution, allowing us to fit
the cumulative density distribution of the data with a function of the
form of Eq.~\ref{cf_f}. Note however that due to this approximation,
the derived equations in this section do not reflect the exact
statistics of the galaxy background. However by tuning the parameters
through simulations (see Section 3.4), the resulting statistical
approximation is adequate for our purposes. 

The aim of the fitting procedure is to calculate the average density
of the background cells, so we can subsequently impose a lower limit
on the density of the cells that are caused by clustering. However, we
can only fit the function to the lower-density end of the distribution
which is not influenced by the cells in the overdense
regions. Therefore we first estimate the background density by
inspecting the histogram of the cell densities.

\begin{figure}
%\vspace{-1.2cm}
\hspace{-0.6cm}
\includegraphics[height=8.3cm]{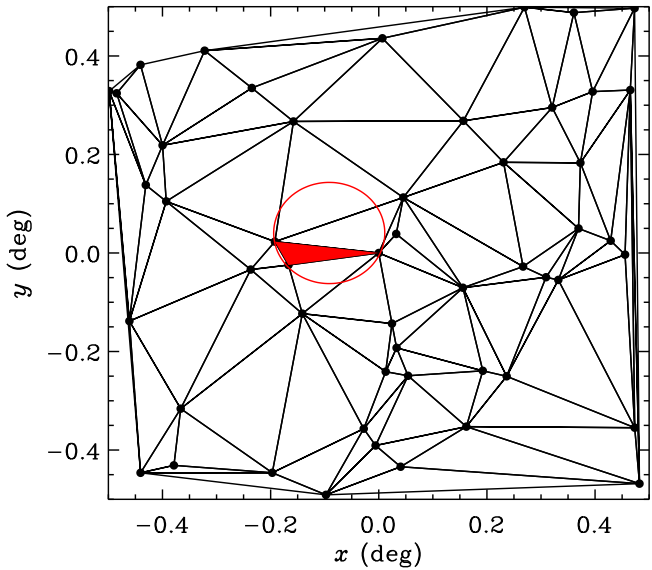}
\vspace{-0.5cm}
\caption{\small An example of Delaunay triangulation (Delaunay
  1934). The dots represent galaxies in the field which is the same as
  in Fig.~\ref{voronoi}. The filled (red) triangle and circle demonstrate the
  definition of the Delaunay Triangulation: the circumcircle of any
  triangle contains no other points than the vertices of the triangle
  itself.}
\label{delaunay}
\end{figure}

Fig.~\ref{vorhist} shows the VT cell density distribution in a field
of 2000 randomly distributed background galaxies, containing a
structure of 100 galaxies in the centre with a Gaussian density
profile with $\sigma = 1\am$ (see also Fig.~\ref{vor_func}). If we
assume the peak in this histogram is not polluted by the overdense
regions, the form of Eq.~\ref{kiang} dictates that the average
background density is $\frac{4}{3}$ times the density at which the
peak occurs. This can be shown by requiring that the derivative of
Eq.~\ref{kiang} is zero and applying Eq.~\ref{tilde_f}. Next we can
fit the predicted cumulative distribution function to the cumulative
distribution function of our data where $\tilde f_{\rm estimated} \leq
0.8$, as suggested by Ebeling \& Wiedenmann (1993). Once the exact
background density is known, we isolate all cells with $\tilde f >
\tilde f_{\rm min}$; $\tilde f_{\rm min}$ is the density at which
overdense regions start to contribute significantly to the cumulative
density distribution. Adjoining high-density cells are grouped
together; if the group consists of a number greater than a certain
lower limit, it is taken to be a cluster
candidate. Fig.~\ref{vor_func} illustrates this procedure: the Voronoi
tessellated field is shown on the left along with the high-density
groups and the cluster candidate; on the right the cumulative density
distribution is plotted. The limiting number of galaxies, $n_{\rm
  lim}$, can be calculated by setting a lower limit to $N_{\rm exp}$:
the expected number of groups caused by background
fluctuations. Ebeling \& Wiedenmann (1993) derived this quantity as
described below in Eqs.~\ref{Nfluct} - \ref{Nexp}. Note that we use
the lower case notation $n$ for numbers of {\it individual} Voronoi
Cells (each representing a galaxy), and the capital $N$ for numbers of
high-density {\it groups} of Voronoi Cells (corresponding to cluster
candidates).

The expected number of groups caused by background fluctuations,
comprising a certain number of galaxies above the background level,
$n_{\rm gal}$, can be written as: \be N_{\rm fluct}(\tilde f_{\rm
min},n_{\rm gal}) = n_{\rm bg} N_{\rm fluct}(\tilde f_{\rm min},0)
e^{-b(\tilde f_{\rm min})n_{\rm gal}},
\label{Nfluct}
\ee where $\tilde f_{\rm min}$ is the minimum density cut-off value
used to select high-density cells, and $n_{\rm bg}$ is the number of
background galaxies expected in the field. The latter comes directly
from the fitted average background density $<\!\!f\!\!>$ by
recognising that $<\!\!a\!\!> = 1~/<\!\!f\!\!>$ and therefore $n_{\rm
  bg} = A~/<\!\!a\!\!>$, where $A$ is the total area of the survey
field. $N_{\rm fluct}(\tilde f_{\rm min},0)$ is the number of
high-density groups with no extra galaxies above the background level;
$N_{\rm fluct}(\tilde f_{\rm min},0)$ and $b$ have been shown by
Ebeling \& Wiedenmann (1993) to obey the following empirical
relations: 
\be
 N_{\rm fluct}(\tilde f_{\rm min},0) = 0.047\tilde f_{\rm min} - 0.04,
\label{nfluct_0}
\ee
\be 
b(\tilde f_{\rm min}) = 0.62 \tilde f_{\rm min} - 0.45.
\label{b}
\ee
Integrating the function given in Eq.~\ref{Nfluct} from the limiting
number of galaxies to infinity gives the expected number of groups
caused by background fluctuations with $n_{\rm gal} > n_{\rm lim}$:
\be
N_{\rm exp}(\tilde f_{\rm min}, n_{\rm gal} > n_{\rm lim}) = n_{\rm
  bg} \frac{N_{\rm fluct}(\tilde f_{\rm min},0)}{b(\tilde f_{\rm
    min})} e^{-b(\tilde f_{\rm min})n_{\rm lim}}.
\label{Nexp}
\ee
Thus, the limiting number of galaxy members in a group considered to
be a cluster candidate is:
\be
n_{\rm lim} = \frac{-1}{b(\tilde f_{\rm min})}\ln\Bigl\{\frac{b(\tilde f_{\rm
      min})N_{\rm exp}}{N_{\rm fluct}(\tilde f_{\rm
      min},0)n_{\rm bg}}\Bigr\}.
\label{nmin}
\ee The number of galaxy members, $n_{\rm gal}$, is determined for
each group and compared to $n_{\rm lim}$. Note that $n_{\rm gal}$
needs to be corrected for the background number density of galaxies,
which is calculated by dividing the total area of the group, $A_{\rm
  group}$, by the average cell area: $n_{\rm gal,bg}~=~A_{\rm
  group}~/<\!\!a\!\!>$. The Voronoi Tessellations method thus has two
parameters for which a value needs to be chosen: the minimum cut-off
dimensionless density $\tilde f_{\rm min}$ and the maximum expected
number of groups caused by background fluctuations, $N_{\rm exp}$.

\begin{figure}
\hspace{-0.6cm}
\includegraphics[height=8.3cm]{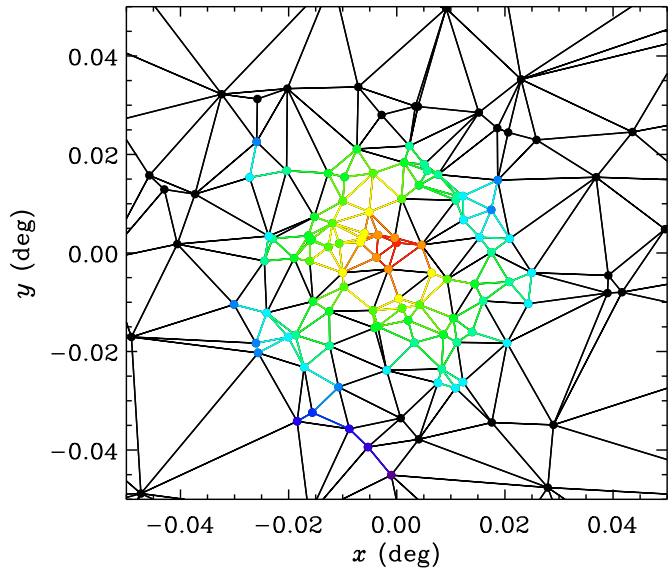}
\vspace{-0.5cm}
\caption{\small The Friends-Of-Friends detection method applied to a
  field of background sources with a central overdensity superimposed,
  exactly as in Fig.~\ref{vor_func} ({\it left}). The background
  consists of 2000 galaxies uniformly distributed throughout a $0.5
  \times 0.5$\,deg$^2$ field; the central structure comprises 100
  galaxies and has a Gaussian density distribution with a $\sigma =
  1\am$. Only the central $0.05 \times 0.05$\,deg$^2$ is shown for
  clarity. The FOF algorithm was run with a linking distance of
  $D_{\rm link} = 175\,\rm kpc$, with the simulated slice being at
  $z=0.5$. The colours of the galaxies and links reflect the iteration
  of the algorithm: the red galaxy was chosen first, the orange ones
  are its `friends', the yellow ones are `friends-of-friends', etc.}
\label{fof}
\end{figure}

\subsubsection{Friends-Of-Friends}\label{FOF}

Friends-Of-Friends algorithms are commonly used in spectroscopic
galaxy surveys (e.g. Tucker et al. 2002; Ramella et al. 2002). A
variant of this algorithm utilising photometric redshifts was proposed
by Botzler et al. (2004). They create redshift slices for their data
cube and place the galaxies into the redshift slices according to their
photometric redshift and error; objects with large errors are
removed. The algorithm then calculates the distance of one galaxy to
all others in the redshift slice, and groups the galaxies that are
closer to each other than a given linking distance, $D_{\rm link}$
(`friends'). Next it calculates the distance from the new galaxies in
the group (the `friends') to all other galaxies in the slice and adds
those that are within the linking distance (`friends-of-friends'). The
group is complete when there are no more galaxies to be found within the
linking distance to any of the group members. If the group comprises a
number of galaxies above a specified minimum number, $n_{\rm min}$, it
is a cluster candidate. Cluster candidates in separate redshift slices
that contain one or more identical galaxy members are linked up as one
and the same cluster candidate. 

Our Friends-Of-Friends algorithm is broadly similar to that of Botzler
et al. (2004). However, we have made three key improvements, which
will be discussed below. 

First, to speed up the computational efficiency, we apply Delaunay
Triangulation (Delaunay, 1934) to the field of galaxies in the
redshift slice to identify each galaxy's nearest neighbours (`Delaunay
neighbours'). This procedure uses the `divide-and-conquer' method
described in Lee \& Schachter (1980), which has a very short
computational run time. Hereby our computation time is greatly reduced
as once we have completed the triangulation, there is no need to
calculate the distance from each galaxy to every other galaxy in the
field, but only to determine the distance to each galaxy's Delaunay
neighbours. Fig.~\ref{delaunay} demonstrates the principle of Delaunay
Triangulation: each galaxy is connected to its nearest neighbours,
forming triangles whose circumcircle contains no other galaxies than
the ones that form the vertices of the triangle itself.

When the triangulation is complete, a random galaxy is chosen and the
proper distance, $D$, to its neighbours as linked by the Delaunay
triangulation, is calculated from:
\be
D = 2\sin\bigl(\frac{\theta}{2}\bigr)D_{\rm A}, 
\label{dist}
\ee
where $D_{\rm A}$ is the angular distance of the redshift slice, and
$\theta$ is the angle between the galaxies $i$ and $j$ in the
tangent-plane approximation:
\be
\theta = \sqrt{\bigl(\alpha_i \cos{(\delta_i)}-
\alpha_j\cos{(\delta_j)}\bigr)^2 + \bigl(\delta_i - \delta_j\bigr)^2}.
\label{theta}
\ee In this equation $\alpha$ and $\delta$ are the RA and Dec of the
galaxies in units of degrees. Any neighbours for which $D \leq D_{\rm
  link}$ are dubbed `friends' and are added to the group. Next, the
previous step is repeated for the new `friends', taking only the
galaxies into account that are not yet members of the group. When
there are no more `Delaunay neighbours' of any members of the group
within linking distance, an as yet unanalysed galaxy is chosen and the
whole process is repeated. This is illustrated by Fig.~\ref{fof},
where the Delaunay Triangulation is shown of a galaxy field with an
overdensity superimposed and the iterations of the Friends-Of-Friends
process are colour-coded. When all groups have been found in the
redshift slice, only those with a number of galaxies greater than
$n_{\rm min}$ are retained. Evidently, the two parameters in FOF for
which a value needs to be chosen are $D_{\rm link}$ and $n_{\rm min}$.

The second important difference between our algorithm and previous
ones in the literature, such as Botzler et al. (2004), is the way we place
the galaxies in the redshift slices. As we sample the full z-PDF to
create MC-realisations of the three-dimensional galaxy distribution,
we do not need to assign errors to individual galaxy redshifts. An
object with a large redshift error will be distributed throughout many
different slices in the 500 MC-realisations, and therefore not yield a
significant contribution to the cluster candidates it is potentially
found in. Thus there is no need to remove objects with large errors
from the catalogue and no additional bias is introduced against faint
objects with noisier photometry.

The third modification to existing algorithms is the way we link up
cluster candidates throughout the redshift slices. Instead of
comparing individual galaxies in the clusters and linking up the
clusters with corresponding members (see Botzler et al. 2004), we use
probability maps of all redshift slices to locate likely cluster
regions. This is discussed in Section~\ref{prob_maps}.

\subsection{Probability maps and cross-checking}\label{prob_maps}

\begin{figure}
\hspace{-0.5cm}
\includegraphics[width=9cm]{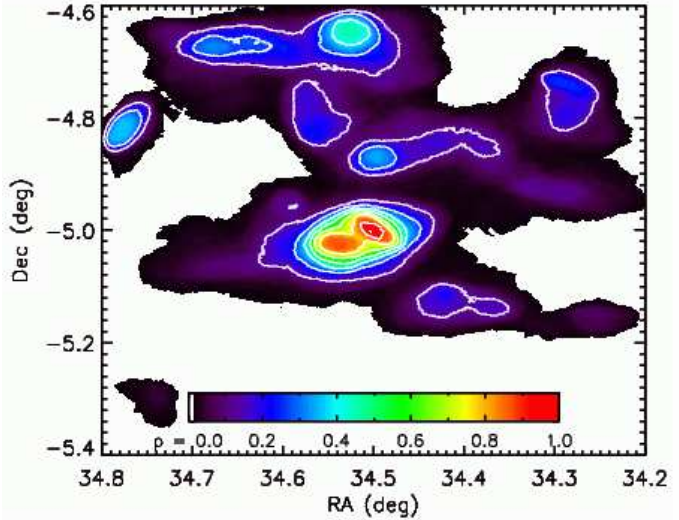}
\vspace{-3mm}
\caption{\small A probability map of clusters found
  by the Voronoi Tessellation method at redshift $z \sim 1.0$. Colours
  are normalised to the highest probability in the field.}
\label{contours}
\end{figure}

Once the two cluster selection methods have determined the cluster
candidates in the redshift slices for all MC-realisations, we combine
the MC-realisations to create probability maps for both methods for
each redshift slice. These maps are created by calculating the extent
of all cluster detections in RA and Dec according to the positions of
the cluster members. The regions of the field that are found to be in
a cluster in many MC-realisations are high-probability cluster
locations. Fig.~\ref{contours} shows an example of a probability map:
the VT cluster candidates in this slice at $z = 1.0$ are contoured and
coloured, with black through to red indicating low to high
probability.

\begin{figure}
%\vspace{2.5mm}
\hspace{-1cm}
\includegraphics[width=9.5cm]{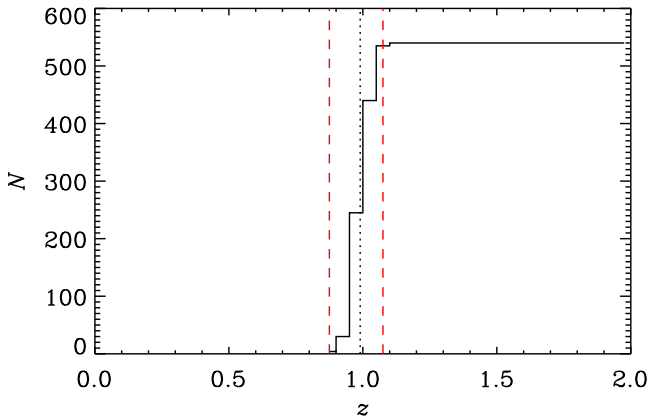}
\vspace{-5mm}
\caption{\small The cumulative number of cluster candidates, which
  form the constituents of one particular cluster, versus
  redshift. The cluster candidates are linked up between the redshift
  boundaries marked by the dashed (red) lines. The final cluster consists of
  $\sim 540$ cluster candidates in different MC-realisations, spread
  out over four redshift slices around $z \sim 1$.  The maximum number
  of constituent candidates would be $2000$, if the cluster was
  detected in all four slices in all MC-realisations. The dotted line
  marks the weighted average redshift of the cluster.}
\label{zlink}
\end{figure}

Since the error on the photometric redshifts of the galaxies is
usually larger than the width of the redshift slices, each cluster
candidate is typically found in several adjoining slices. We join the
cluster candidates that occur in the same location in several slices
by locating the peaks in the probability maps and inspecting the area
within their contours in the adjoining redshift slices for cluster
candidates. This procedure is carried out as follows: per redshift
slice, starting at the highest detected contour level, we calculate
the positions of the cluster contours and determine their 'centres of
mass', where each point within the contour is assigned
an equal 'mass'. Next, we inspect the contours one level down, and
verify if any of these are unoccupied by any of the previously found
centres. If so, this is labelled a new cluster (of a lower
probability). We continue until we have inspected all contour levels
down to 0.05 (or 5\% of the number of MC realisations) in all redshift
slices. Finally, we join each cluster centre to the cluster centres in
adjoining redshift slices that lie within 0.5 Mpc in projected
distance. Fig.~\ref{zlink} shows the cumulative number of
MC-realisations versus redshift for one cluster candidate. The
redshift limits of the linking procedure are placed at the slices
where the cluster candidate is no longer found in a significant number
of MC-realisations (i.e. $< 2.5\%$ of the MC realisation). The final
cluster redshift is determined by taking the mean of the redshift
slices, weighted by the number of MC-realisations in which the
candidate is detected.

We assign a reliability factor $F$ to each cluster by counting the
total number of MC-realisations in which it occurs in any of the
linked-up redshift slices, and dividing this by the total of 500
realisations. This means that if a cluster candidate occurs in four
slices in a single realisation, it is only counted once. Therefore the
maximum number of realisations in which it is counted is 500, in which
case we would have $F = 1.0$. To create the final cluster catalogue,
we cross-check the output of the two detection methods and select only
those clusters that have been found by both VT and FOF with a
reliability factor $F$ above a suitable limit. This parameter $F_{\rm
  lim}$ is dependent on the accuracy of the photometric redshifts, and
the completeness and efficiency of both detection methods. The higher
the chosen limit, the more efficient yet the less complete the final
cluster catalogue will be. The best level of $F_{\rm lim}$ is
determined by simulating mock catalogues, taking into account the
characteristics of the data to be used. Below we describe the results
of running the 2TecX algorithm on our simulated catalogues. Based on
these, VB06 used a value of $F_{\rm lim} = 0.2$ to obtain a reliable
cluster catalogue at $0.5 < z < 1.5$.

\section{Simulations}\label{sim}

\subsection{Mock catalogue characteristics}
To test the behaviour of the cluster-detection algorithm and to
determine the optimal values of the parameters we run a set of
simulations on mock catalogues. These catalogues need to mimic as
closely as possible the data to which the algorithm will be
applied. VB06 describe the application of our algorithm to a combined
optical/infrared catalogue on the Subaru-{\it XMM-Newton} Deep Field
(SXDF) consisting of $BVRi^{\prime}z^{\prime}$ Subaru SuprimeCam data;
$JK$ United Kingdom InfraRed Telescope (UKIRT) Wide Field CAMera
(WFCAM) data from the UKIRT Infrared Deep Sky Survey (UKIDSS); and 3.6
and 4.5 $\mu m$ bands data from the {\it Spitzer} InfraRed Array
Camera (IRAC). Our mock catalogues are designed to have the same area
and $K$-band limiting magnitude as the data catalogue of
VB06. Furthermore, when a galaxy's z-PDF is needed, this is randomly
drawn from the collection of z-PDFs used by VB06 that peak at the
position of the simulated galaxy's redshift. Thus the z-PDFs of the
simulated data accurately reflect the photometric redshift error and
the functional form of the z-PDFs in the real data catalogue.

\begin{figure}
%\vspace{-5mm}
\hspace{-0.7cm}
\includegraphics[width=9.5cm]{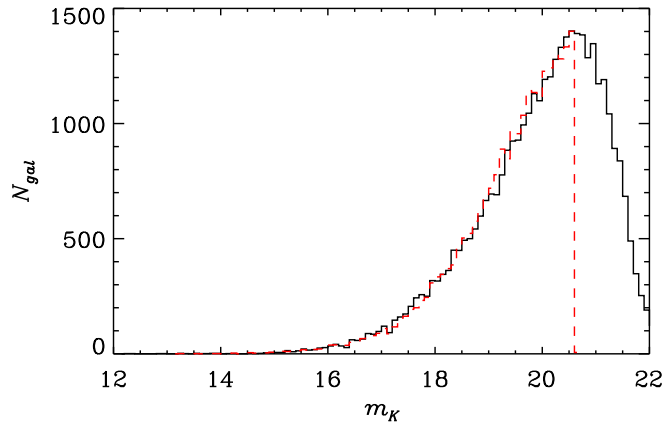}
\vspace{-8mm}
\caption{Number counts versus magnitude for VB06's data catalogue
  (black) compared to the number counts of the mock background
  catalogue (dashed, red). The 5-$\sigma$ detection limit is $K_{\rm lim} =
  20.6$.}
\label{numbercounts}
\end{figure}

\subsection{Simulating the galaxy background}
We create catalogues with a galaxy background distribution randomly
placed in the field with $0.1 \leq z \leq 2.0$ (neglecting clustering
of both the background and the clusters). The galaxy luminosities and
number densities are determined by the $K$-band Schechter luminosity
function of Cole et al. (2001) with $\Phi^* = 3.7\times 10^{-3}\,\rm
Mpc^{-3}$, $\alpha = -0.95$, and $M_K^* = -24.18$. To obtain the
correct value for $M_K^*$ we added $0.017$ (Hewett et al. 2006) to
Cole's original value to account for the difference between the
$K$-band filters of WFCAM and 2MASS (used by Cole et al. 2001). Also,
we assume passive evolution of the luminosity function (e.g. Gardner
et al. 1996). We calculate the $e$+$k$ (evolution and redshifting)
correction to $M_K^*$ at all redshifts by using {\it GALAXEV} to
create a stellar population synthesis SED. The SED consists of a
star-burst at $z=4$, exponentially decaying with $\tau = 1 \rm\, Gyr$,
and has solar metallicity. The creation of the background catalogues
is done in the following steps:
\begin{enumerate}
\item We slice the three-dimensional field into redshift slices of
   $\Delta z = 0.05$ over which we assume the luminosity function to
   be constant.
\item For each slice, we calculate the volume ($V$), determined by the
  angular size of the field and the redshift limits, and the
  $e$+$k$ corrected $M_K^*$.
\item The number of simulated galaxies in the slice is calculated according to
  the luminosity function:
\be
N_{\rm gal} = V \times \int_0^\infty\Phi(L)dL, 
\label{ngal}
\ee
and 
\be
\Phi(L)dL = \Phi^*{\mathcal L}^\alpha e^{-{\mathcal L}}d{\mathcal L}, 
\label{phi}
\ee
where ${\mathcal L}=L/L^*$ is a dimensionless luminosity.
\item Luminosities are assigned to all galaxies according to the
  luminosity function of Eq.~\ref{phi}, and the absolute magnitudes are
  determined with:
\be
M_K = M_K^* - 2.5 \log{\mathcal L}.
\label{mabs}
\ee
\item The galaxies are randomly placed in redshift, RA, and Dec within
  the slice according to a uniform distribution. 
\item The apparent magnitudes of the simulated galaxies are
  calculated:
\be
m_K = M_K + 5\log(D_{\rm L})-5, 
\label{mapp}
\ee where $D_{\rm L}$ is the luminosity distance to the galaxy in
parsec. We now impose a magnitude limit of $K_{\rm lim} < 20.6$ to
match the 5-$\sigma$ limit of the data catalogue of VB06. Only the
galaxies with $m_K < K_{\rm lim}$ are retained in the mock catalogue.
\end{enumerate}
The number of galaxies as a function of magnitude in each mock
catalogue is entirely consistent with the number counts in the data
catalogue up to the 5-$\sigma$ limit, as is shown in
Fig.~\ref{numbercounts}.

\begin{figure*}
\hspace{-1cm}
\includegraphics[width=18cm]{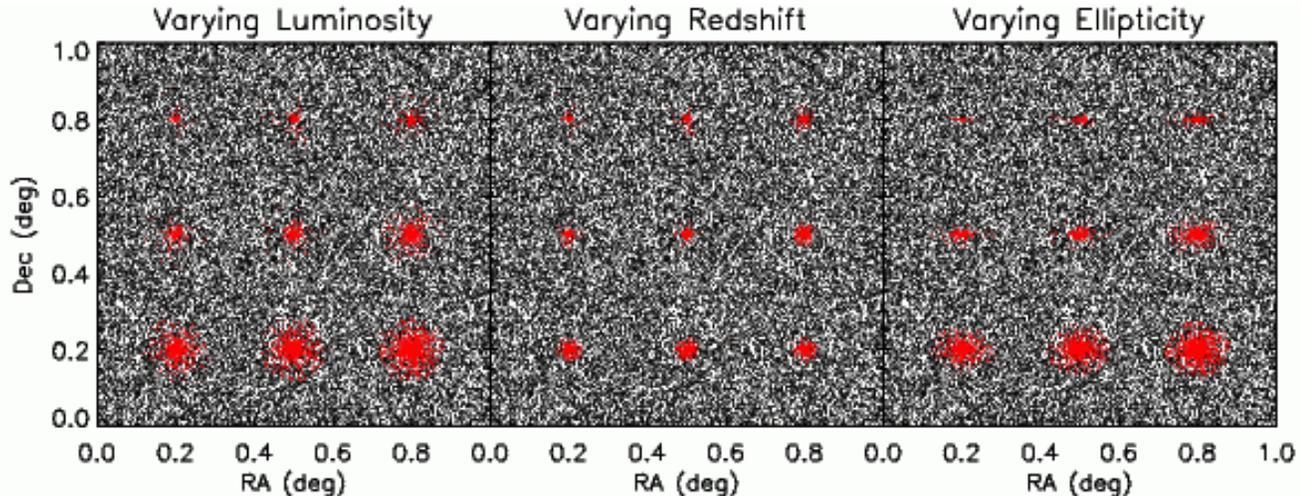}
\vspace{-5mm}
\caption{\small Example of simulated equatorial fields containing
  three types of clusters (red) superimposed on a galaxy background
  (black) at $0.1 < z < 2.0$. {\it Left:} Nine clusters at $z = 0.2$
  with total luminosities of $L_{\rm tot}$ = 10, 20, 30, 40, 50, 100,
  150, 200, 300 $L^*$. {\it Middle:} Nine clusters of $L_{\rm tot} =
  50\,L^*$ at $z$ = 0.2, 0.4, ..., 2.0. {\it Right:} Nine clusters of
  $L_{\rm tot} = 50\,L^*$ at $z = 0.2$, with ellipticity $e$ = 0.1,
  0.2, ..., 1.0 (PA = 0$\deg$).}
\label{mock_cat}
\vspace{-4mm}
\end{figure*}

\subsection{Adding mock clusters to the catalogue}

 We superimpose simulated clusters on the background catalogue. To
 create the mock clusters we take the following steps:
\begin{enumerate}
\item We choose a total cluster mass (including dark matter) and a
  mass-to-light ratio of $M{\rm\,[M_{\odot}]} / L{\rm\,[ L_{\odot}]} =
  75h$ (Rines et al. 2001) which is assumed constant {\it in terms of
    L$^*$} (a quantity we assume to evolve passively with
  redshift). To deduce the total luminosity of the cluster in $K$-band
  we calculate: 
\be L^* = 10^{\bigl(\frac{K_\odot-M_K^*}{2.5}\bigr)}\rm\,
  L_{\odot},
\label{lstar}
\ee
and therefore the total dimensionless luminosity in units of $L^*$ is:
\be
 {\mathcal L}_{\rm tot} = M_{\rm tot}{\rm [M_\odot]}/\Bigl(75h
     10^{\bigl(\frac{K_\odot - M_K^*}{2.5}\bigr)}\Bigr).
\label{ltot1}
\ee Here $K_\odot = 3.28$ is the $K$-band magnitude of the sun and
$M_K^*$ is taken from the cluster luminosity function derived by Lin,
Mohr \& Stanford (2004), who found $M_K^* = -24.34$, $\Phi^* = 3.0
\rm\, Mpc^{-3}$, and $\alpha = -1.1$. Again we assume passive
evolution of the cluster luminosity function with a formation redshift
of $z_{\rm form} = 4$.
\item We calculate the number of galaxies in the cluster by using
  Eq.~\ref{ngal} and recognising that:
\be
L_{\rm tot} = V \times \int_0^\infty L\Phi(L)dL.
\label{ltot2}
\ee
Together this gives:
\be
N_{\rm gal} = L_{\rm tot} \times
\frac{\int_0^\infty\Phi(L)dL}{\int_0^\infty L\Phi(L)dL},
\label{ngal2}
\ee
or in units of $L^*$:
\be
N_{\rm gal} = {\mathcal L}_{\rm tot} \times
\frac{\int_0^\infty{\mathcal L}^\alpha e^{-{\mathcal
      L}}d{\mathcal L}}{\int_0^\infty{\mathcal L}^{\alpha+1}e^{-{\mathcal
      L}}d{\mathcal L}}.
\label{ngal3}
\ee
Luminosities are assigned to the galaxies according to the luminosity
function of Lin et al. (2004).
\item The galaxies are spatially distributed within the cluster
  according to an NFW profile (Navarro, Frenk \& White, 1997) with a
  cut-off radius of 5\,Mpc. Assuming galaxies to be perfect tracers of
  the dark matter, the galaxy number density $n$ in the two-dimensional
  projected NFW profile is (Bartelmann 1996): 
\be 
n = \left\{
\begin{array}{lr}
\frac{1}{x^2-1} \Bigl( 1 -
 \frac{\ln\frac{(1+\sqrt{1-x^2})}{x}}{\sqrt{1-x^2}}\Bigr) & x<1 \\
\frac{1}{3} & x = 1 \\
\frac{1}{x^2-1} \Bigl( 1 -
 \frac{\arctan\sqrt{x^2-1}}{\sqrt{x^2-1}} & x>1 
\end{array}
\right.
\label{nfw}
\ee Here $x=r/r_s$, where $r$ is the radius in projection. The scale radius $r_s$ is
related to $r_{200}$ (the radius of the circle whose density is 200
times the critical density of the Universe) via $c = r_{200} /
r_s$. The concentration factor $c$ has been determined from numerical
simulations by Dolag et al. (2004) to obey the empirical relation: 
\be (1+z)c = c_{\rm 0} \Bigl(\frac{M}{M_0}
\Bigr) ^ \alpha,
\label{conc}
\ee
with $c_{\rm 0} = 9.59$, $M_{\rm 0} = 10^{14}h^{-1}\rm M_{\odot}$,
and $\alpha = -0.1$.\\
The radius $r_{200}$ is determined by the total mass of the cluster:
\be
r_{200} = \sqrt[3]{\frac{M_{200}}{\frac{4}{3}\pi200\rho_{\rm cr}}},
\label{r200}
\ee
where for a flat Universe:
\be
\rho_{\rm cr} = \frac{3}{8\pi G} H_{\rm 0}^2 \Bigl ((1+z)^3\Omega_{\rm
  M}+\Omega_\Lambda \Bigr).
\label{rho_cr}
\ee 
In this equation $H_{\rm 0}$ is expressed in $\rm
km\,s^{-1}\,km^{-1}$, and $G$ is the gravitational constant. When
simulating elliptical clusters, we use the radius $r_{200}$ for the
profile over one axis, and the radius $e \times r_{200}$ over the
other axis, where $e$ is the ellipticity expressed in minor axis over
major axis.
\begin{figure*}
\begin{center}
\includegraphics[width=15cm]{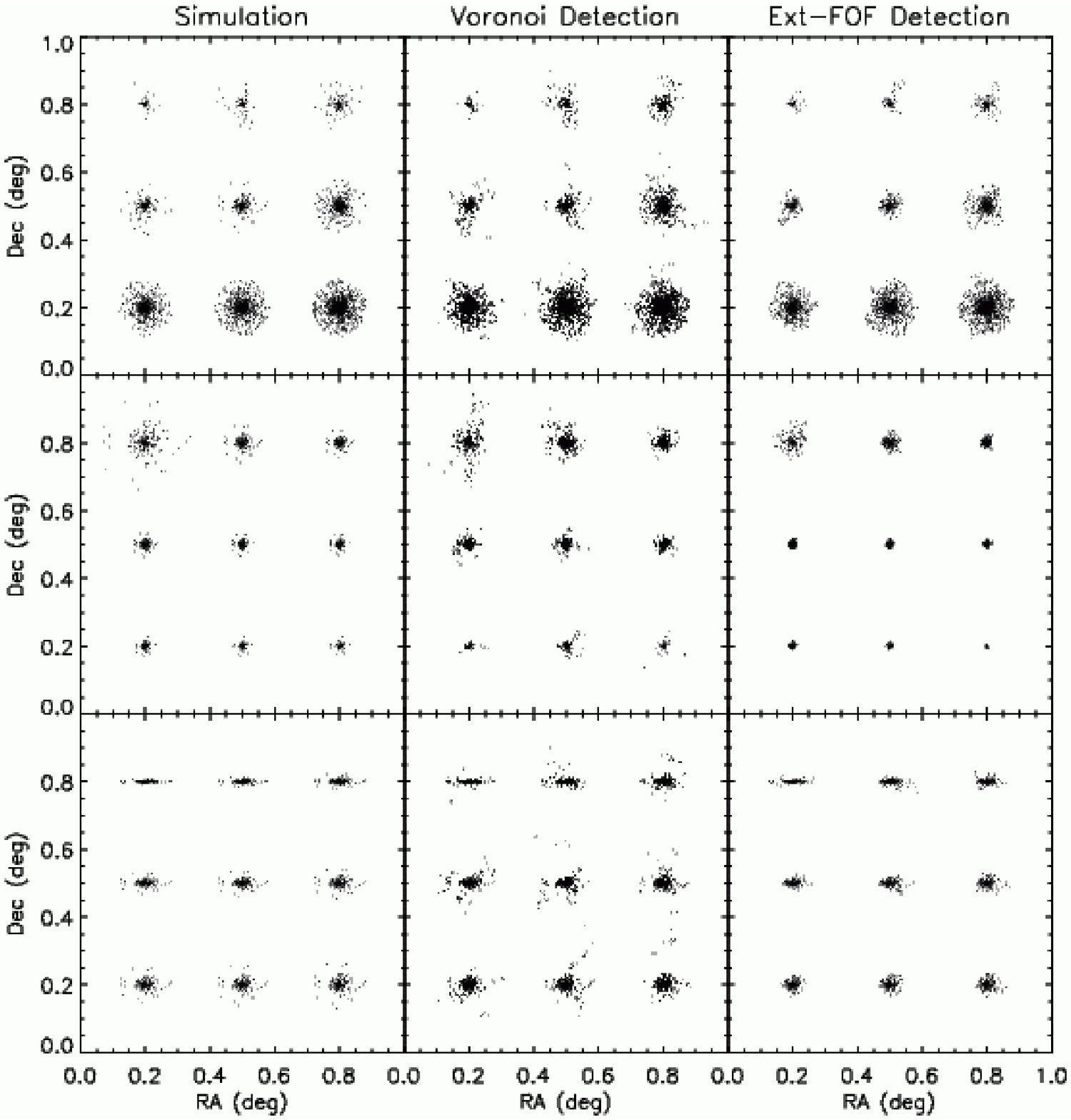}
\vspace{-3mm}
\caption{\small Mock clusters as recovered by the Voronoi
  Tessellations and Friends-Of-Friends algorithms. The left panel
  shows the distribution of cluster galaxies in the mock catalogues;
  note that the background galaxies have been removed from the plot
  for clarity. The middle panel shows the clusters as recovered by the
  VT method, whereas the right panel shows the clusters as recovered
  by the FOF method. The three simulated cluster fields (top to
  bottom) are identical to the ones in Fig.~\ref{mock_cat}, where the
  top panel contains the clusters of varying mass, the middle panel
  the clusters at varying redshift, and the bottom panel the clusters
  of varying ellipticity.}
\label{simulations}
\end{center}
\end{figure*}

\item The redshifts of the cluster galaxies are randomly offset from
  the cluster redshift according to a Gaussian distribution with
  $\sigma = 0.05(1+z)$, which is the expected photometric redshift
  error (see VB06). This error is much larger than the contribution of
  the velocities of the galaxies within the cluster, which allows us
  to neglect the latter.
\item Again, we apply the magnitude limit of $K_{\rm lim} < 20.6$ to
  the apparent magnitudes of the cluster galaxies to obtain the final
  catalogue.
\end{enumerate}

We create different types of mock cluster catalogues: (i) a set of
clusters with varying mass or total luminosity at fixed redshift, (ii)
a set of clusters of fixed mass at varying redshifts, and (iii) a set
of clusters of fixed mass and redshift, but with varying
ellipticity. The varying mass and redshift catalogues are created such
that each combination of mass and redshift is represented and all
catalogues are recreated randomly ten times. In Fig.~\ref{mock_cat} we
show the distribution of galaxies in our three types of catalogue: on
the left nine clusters at $z = 0.2$ with total luminosities of $L_{\rm
  tot}$ = 10, 20, 30, 40, 50, 100, 150, 200, 300 $L^*$; in the middle
nine clusters of $L_{\rm tot} = 50\,L^*$ at $z$ = 0.2, 0.4, ..., 2.0;
on the right nine clusters of $L_{\rm tot} = 50\,L^*$ at $z = 0.2$,
with ellipticity $e$ = 0.1, 0.2, ..., 1.0 at a position angle (PA) of 0$\deg$.

\subsection{Simulation results}\label{results2}

\begin{figure*}
\begin{center}
\includegraphics[width=15cm]{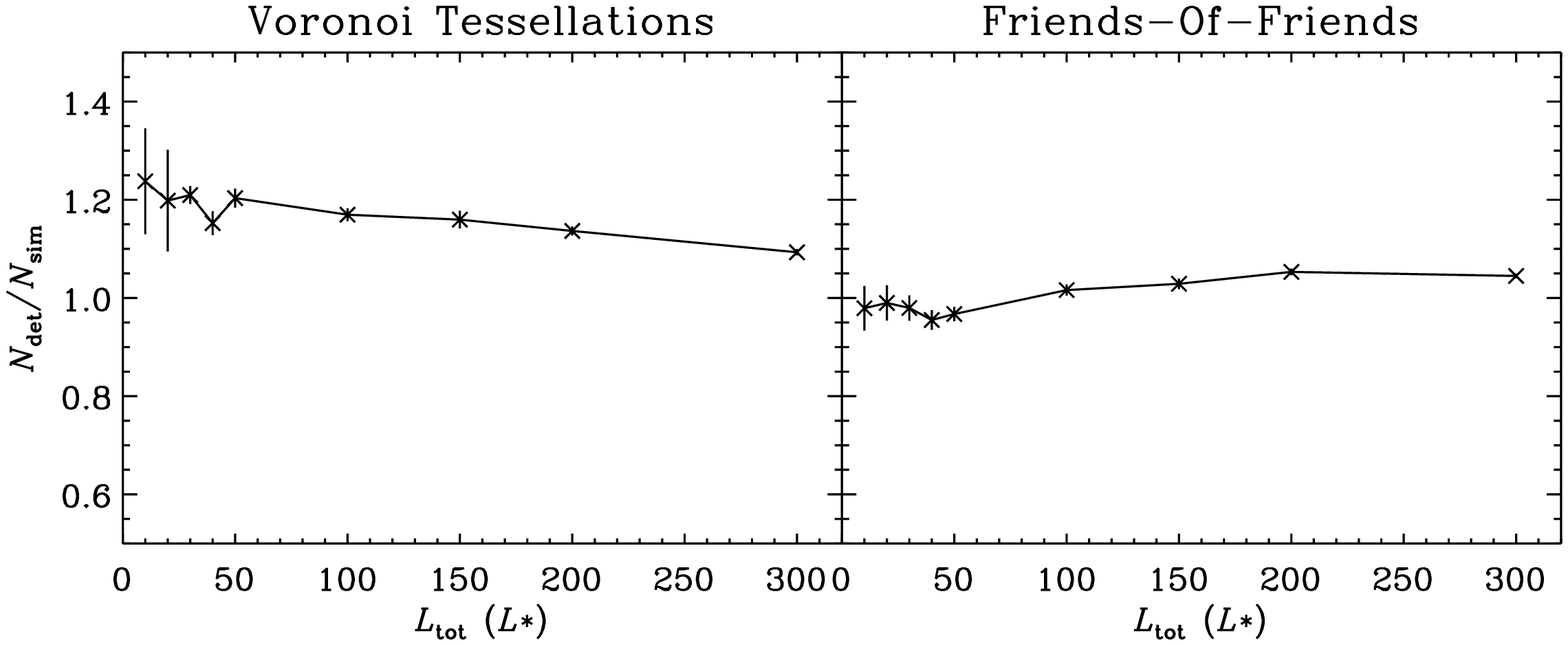}
\includegraphics[width=15cm]{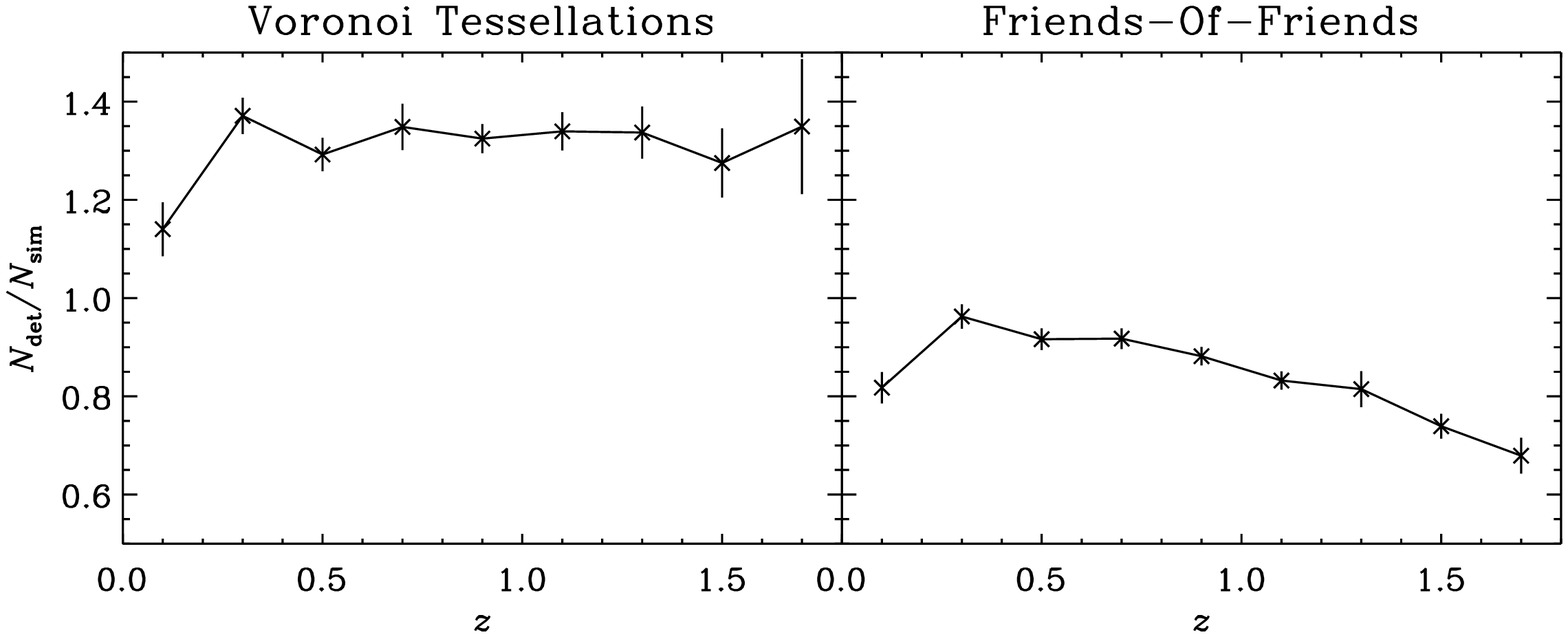}
\includegraphics[width=15cm]{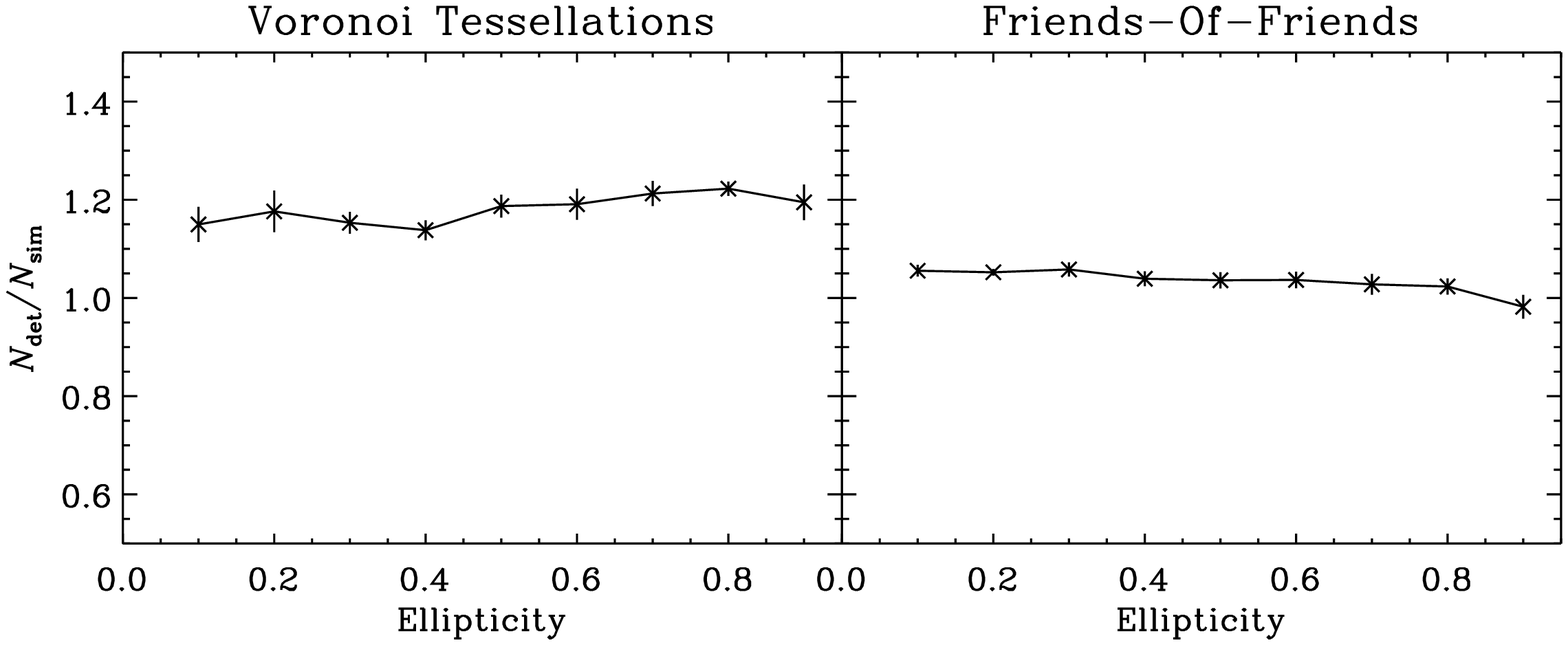}
\caption{\small The fraction of simulated cluster galaxies recovered
  by both detection methods. The number of simulated cluster galaxies
  is determined by the mass of the cluster as well as the magnitude
  limit at the cluster's respective redshift. VT systematically
  overestimates the number of galaxies, whereas FOF recovers a more
  accurate number of galaxies. The top panel shows the recovered
  fraction as a function of cluster luminosity, the middle panel as a
  function of redshift, and the bottom panel as a function of
  ellipticity.}
\label{ngalplot}
\end{center}
\end{figure*}

The aim of the simulations is to explore the behaviour of the FOF and
VT detection methods, and to optimise the algorithm's parameters. The
VT and FOF methods each have two free parameters. For FOF these are
the linking distance in proper coordinates, $D_{\rm link}$, and the
minimum number of galaxies in a cluster, $n_{\rm min}$. Guided by
Botzler et al. (2004) we experimented with values between 0.125 Mpc
$\leq D_{\rm link} \leq$ 0.175 Mpc, and $3 \leq n_{\rm min} \leq
5$. For VT the parameters are the expected number of groups due to
background fluctuations, $N_{\rm exp}$, and the lower limit on the
cell density, $\tilde f_{\rm min}$. We followed the method of Ebeling \&
Wiedenmann (1993) and set $N_{\rm exp}$ to 0.1. For $\tilde f_{\rm
  min}$ we tried values of $1.2-2.2$, where $\tilde f = 1.0$ equates to
the mean cell density of the field. We use the parameters that give
the best completeness of detected clusters whilst keeping the
contamination low: $D_{\rm link} = 0.175$ Mpc, $n_{\rm min} = 5$, and
$\tilde f_{\rm min} = 1.74$.

Now that we have determined each algorithm's optimal parameters, we test the
behaviour of the cluster detection routine by trying to recover the
clusters of the three different types of mock catalogues described in
the previous section. Fig.~\ref{simulations} shows an example: the
left panel contains the simulated clusters, the middle panel the
clusters recovered by VT, and the right panel the clusters recovered
by FOF. Note that in the left panel the background galaxies have been
removed for clarity; naturally they were present when running the
cluster detection algorithm.

\begin{figure*}
\hspace{-1cm}
\includegraphics[width=18cm]{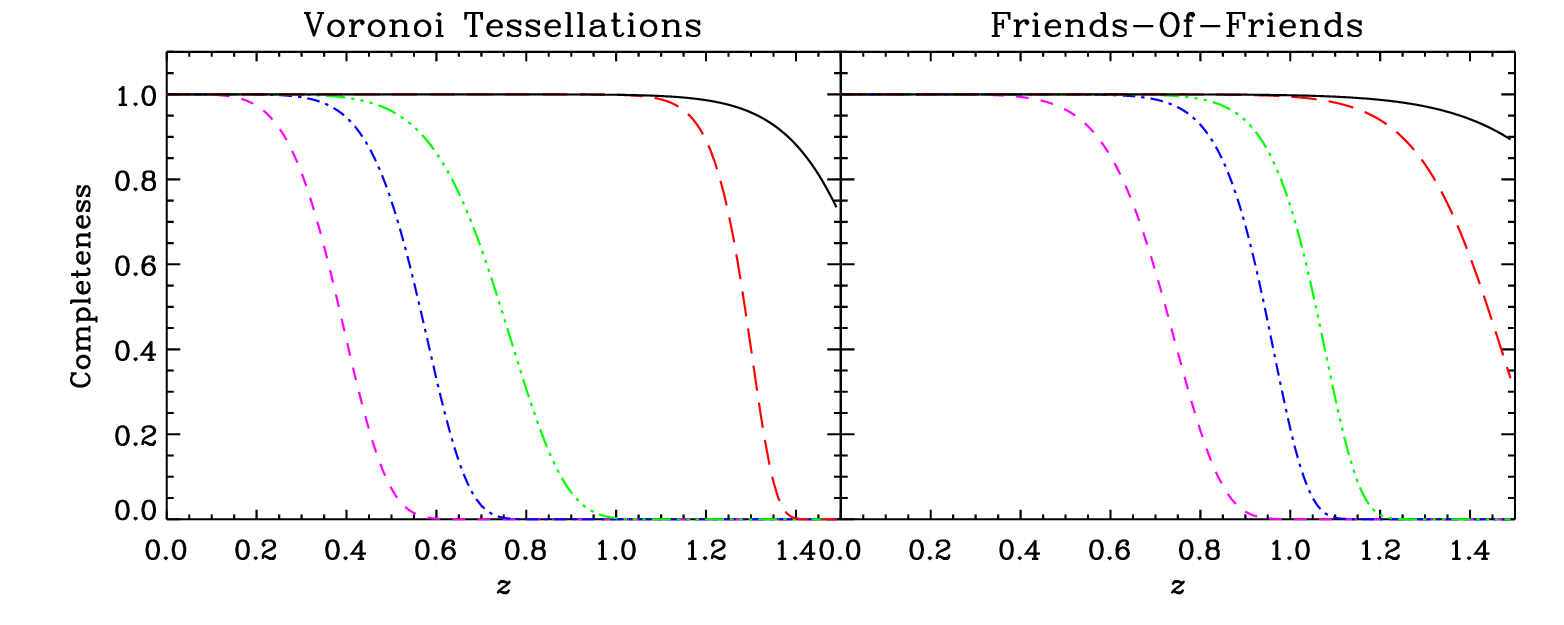}
\vspace{-3mm}
\caption{\small The completeness for different total cluster masses
  at $0 < z < 1.5$. The lines plotted are for total cluster mass
  of 0.5 (purple), 1.0 (blue), 2.0 (green), 10 (red), and $20 \times
  10^{14}\, \rm M_{\odot}$ (black).}
\label{completeness}
\end{figure*}

Both methods recover all clusters satisfyingly; there is no obvious
bias to cluster morphology as the elliptical clusters are recovered
very well by both methods. However, the recovered shape of the
clusters differs for both methods: VT tends to pick up more background
galaxies at the edges of the clusters as the number of recovered
cluster members, $N_{\rm gal}$, in any cluster is sensitive to the
local field density. By contrast, the galaxy members recovered by FOF
are more centrally concentrated; the total number of recovered
galaxies per cluster is consistent throughout the random realisations
of the catalogues. This is illustrated in Fig.~\ref{ngalplot} which
shows the fraction of recovered cluster galaxies by both methods for
the types of catalogues shown in Fig.~\ref{simulations}. The number of
simulated cluster galaxies is determined both by the cluster's mass
and the magnitude limit at its respective redshift. The difference in
both methods is particularly noticeable in the middle panel of
Fig.~\ref{ngalplot}, where the recovered fraction of cluster galaxies
is shown versus redshift. As there are few background galaxies in the
high-redshift slices, the fraction of detected galaxies per cluster
declines in the FOF method as there is a smaller chance of finding
background galaxies within the linking distance. However, the fraction
of detected cluster galaxies remains constant in VT because the
algorithm's parameters to estimate an overdensity are scaled to the
background density, which negates the effect of having less background
galaxies in the redshift slice.

With the chosen set of parameters we can calculate the detection
completeness as a function of redshift for clusters of varying total
mass. Fig.~\ref{completeness} shows the result: clusters of mass
$M_{\rm tot} \sim 2 \times 10^{15} \rm \, M_\odot$ are detected with a
high completeness up to $z = 1.5$, whereas the lower-mass clusters
show rapidly declining completeness at lower redshifts. The FOF
algorithm achieves a higher completeness than VT for clusters of equal
mass; however the contamination of spurious sources is found to be
higher. 

\begin{figure}
%\hspace{-1cm}
\includegraphics[width=8cm,height=6.5cm]{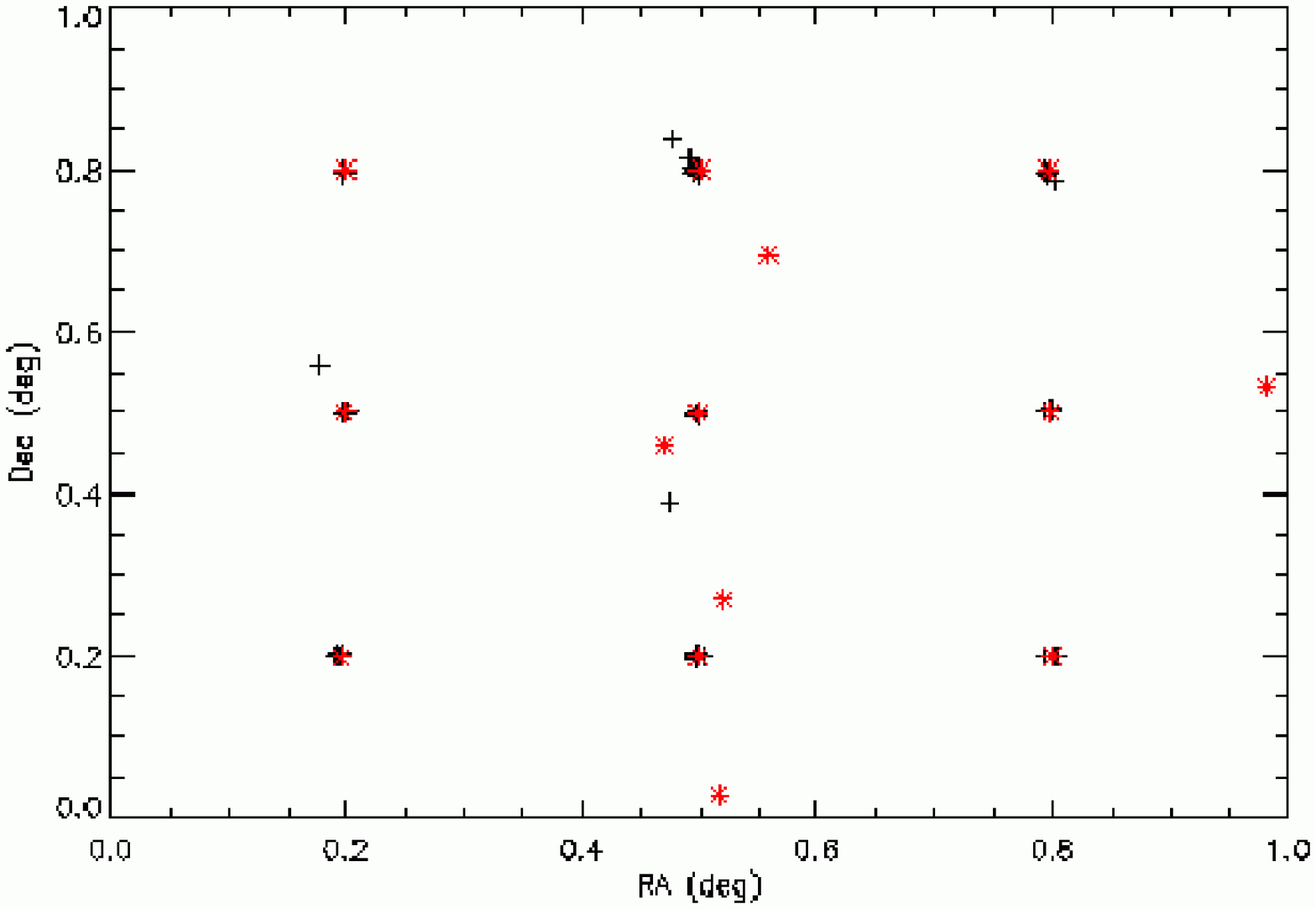}
\vspace{-3mm}
\caption{\small An example of cluster candidates selected in a
  simulated catalogue by both detection methods with $F \ge 0.2$. The
  candidates found by VT are shown in black, the ones found by FOF in
  red. Although both methods detect spurious sources, none are found
  by both.}
\label{detections}
\end{figure}

\begin{figure}
%\hspace{-1cm}
\includegraphics[width=8cm,height=6.5cm]{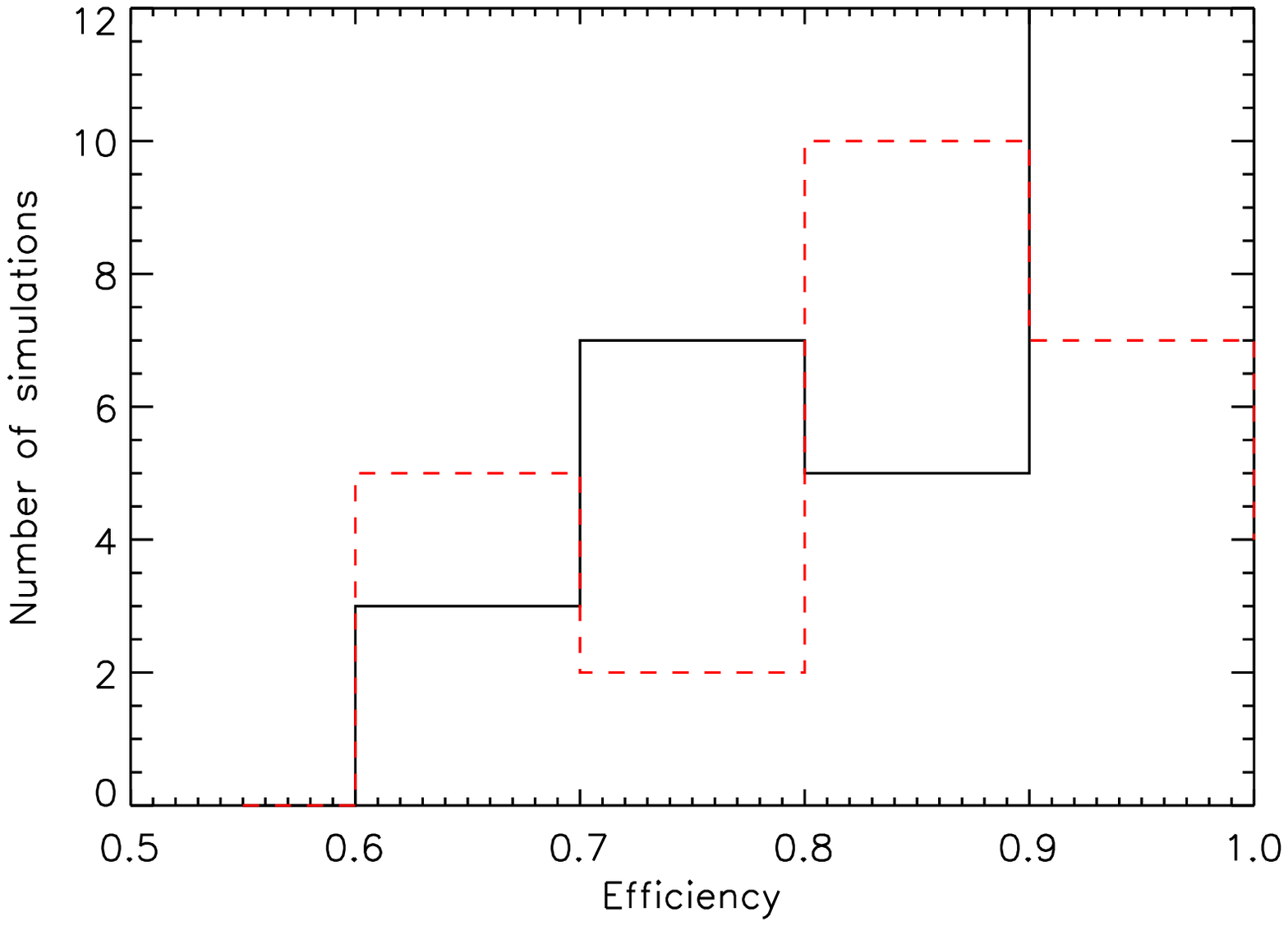}
\vspace{-3mm}
\caption{\small The efficiency, in terms of the number of real clusters as a fraction
  of the total detected clusters, for either method in all 30 mock
  catalogues (black solid line and red dashed line for VT and FOF
  respectively). All clusters found with $F \ge 0.2$ are included in
  this diagram.}
\label{efficiency}
\end{figure}

The effects of contamination of the individual detection methods can
be greatly reduced by cross-checking the output of both methods. Since
both methods use different measures to isolate clusters (galaxy
density in VT versus separation in FOF) the false detections in both
do not typically coincide. Therefore by cross-checking the output of
the two methods and choosing a sensible lower limit for the
reliability factor $F$, the spurious sources due to biases in the
algorithms disappear, leaving only chance galaxy
groupings. Fig.~\ref{detections} is an example of this: it shows the
cluster candidates found in all redshift slices by both methods;
although there are spurious detections both from VT and FOF, none are
found by both. In Fig.~\ref{efficiency} the efficiency, in terms of
the number of real clusters as a fraction of the total detected
clusters, in all 30 mock catalogues is plotted for either method. Here
all clusters with $F \ge 0.2$ are included. The median efficiency is
0.8 for both methods; none of the spurious sources are detected by
both techniques. Note however that this is purely an upper limit to
the efficiency: for a true estimate the proper spatial correlation
function of both background galaxies and clusters need to be taken
into account (for an in-depth discussion of the efficiency for varying
cluster mass and redshift in an accurate spatial model, see the
follow-up paper [Van Breukelen et al. in preparation]). Furthermore,
the quality of the photometric redshifts plays an important role. As
discussed in VB06 and shown in Van Breukelen et al. 2009, artifacts
like redshifts spike can yield a significant number of spurious
sources in the cluster catalogue.

\begin{figure}
\hspace{-1cm}
\includegraphics[width=10cm]{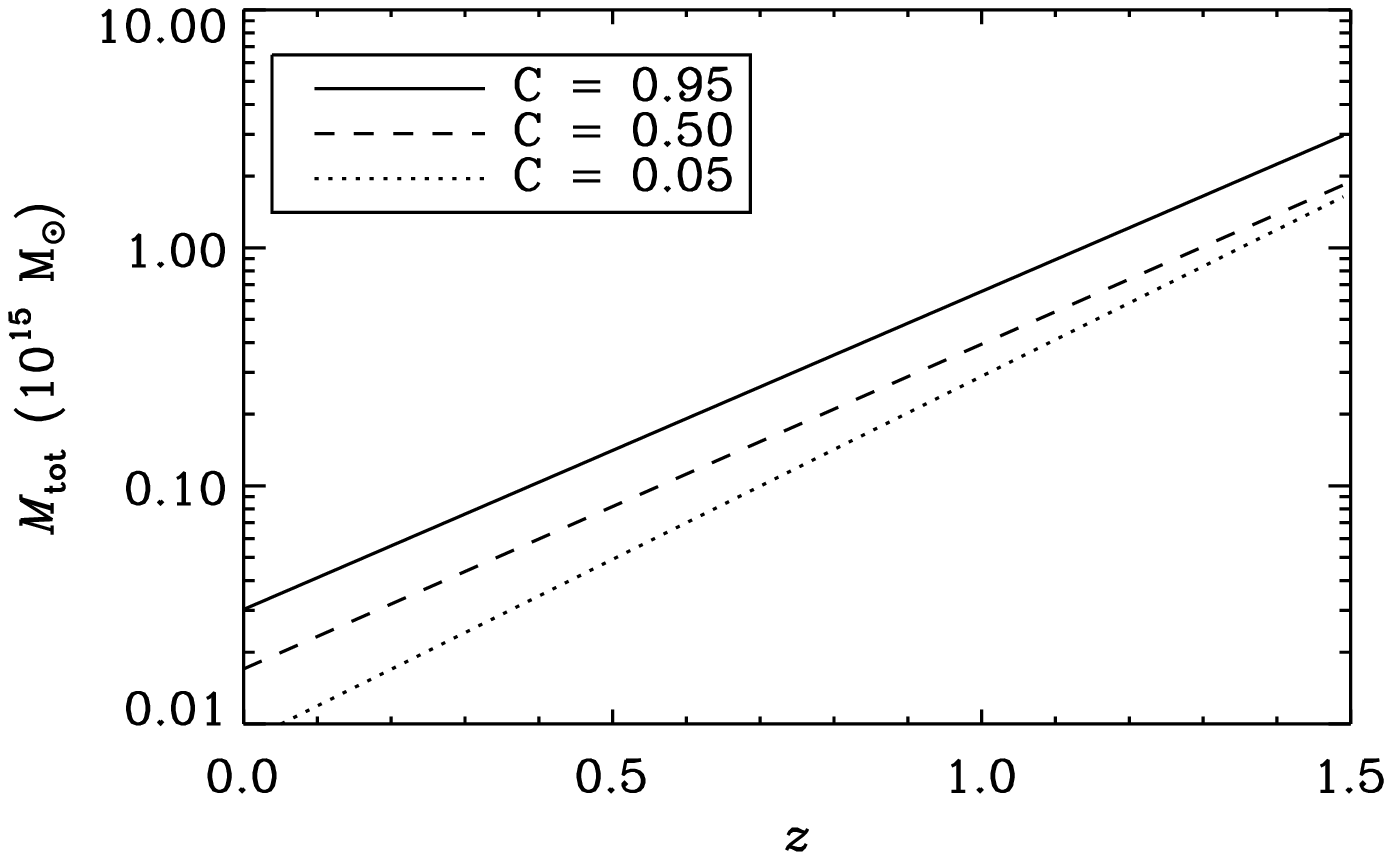}
\vspace{-3mm}
\caption{\small Cluster mass selection function versus redshift. The
  selection function is shown for three completeness (C) levels: 95\%
  (solid line), 50\% (dashed line), and 5\% (dotted line). \bigskip}
\label{selfunc}
\end{figure}

Cross-checking the results of the two methods means the completeness is
limited to the lower value found of the two. With our chosen set of
parameters and keeping only the structures found with $F \ge
0.2$ in both methods, we can calculate the mass selection function of our
algorithm. This is shown in Fig.~\ref{selfunc} for three levels of
completeness.

\begin{figure}
\hspace{-1cm}
\includegraphics[width=10cm]{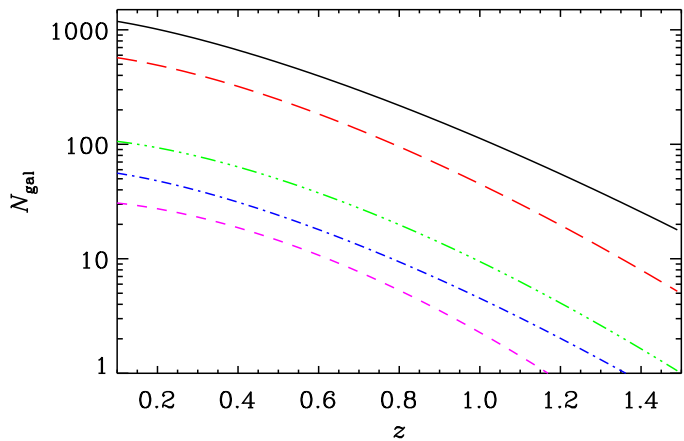}
\vspace{-3mm}
\caption{\small Constant-mass functions for the number of recovered
  cluster members with redshift. The functions plotted are for total
  cluster mass of 0.5 (purple), 1.0 (blue), 2.0 (green), 10, and $20
  \times 10^{14}\, \rm M_{\odot}$ (black). \bigskip}
\label{nz}
\end{figure}

The final application of our simulations is to derive a relationship
between the total cluster mass (or luminosity) and the number of
recovered cluster galaxies. As Fig.~\ref{ngalplot} shows, the number
of galaxies found by FOF is much more consistent and better-behaved
than the number of galaxies detected by VT. Therefore we only use the
FOF output to determine the total cluster mass. This is done by taking
all galaxies that occur in the cluster in $>15\%$ of the
MC-realisations in which the cluster itself is detected. The galaxies
that appear in a smaller fraction of MC-realisations are very likely
to be interlopers from different redshifts.  Calculating $N_{\rm gal}$
for all cluster-masses at all redshifts yields functions of $N_{\rm
  gal}$ vs. $z$ for total constant mass or luminosity. These are shown
in Fig.~\ref{nz}. The number of detected galaxies at constant mass
declines more steeply than a magnitude selected sample would, since the
fraction of recovered versus simulated galaxies for the FOF method
becomes smaller at higher redshift (see Fig.~\ref{ngalplot}).  The
total cluster mass of cluster candidates found in real data (see VB06)
can be estimated by overplotting the number of cluster galaxies and
interpolating between the lines of constant cluster mass.

\section{Summary}
To summarise, the main points of this paper are set out below.

We have created a new cluster detection algorithm of which the main
characteristics are: (i) each galaxy's full redshift probability
function is utilised, and (ii) cluster candidates are selected by
cross-checking the results of two substantially different selection
techniques: Voronoi Tessellations and Friends-Of-Friends.

Each selection technique is dependent on two parameters. Voronoi
  Tessellations uses $\tilde f_{\rm min}$, the limiting cell density,
  and $N_{\rm exp}$, the maximum expected number of groups caused by
  background fluctuations. The parameters of the Friends-Of-Friends
  algorithm are $D_{\rm link}$, the linking distance, and $n_{\rm
    min}$, the minimum number of galaxies in a group.

Simulations using mock background galaxy catalogues with
  clusters superimposed allow us to choose optimum values for the
  algorithm's parameters. We use $N_{\rm exp} = 0.1$, $\tilde f_{\rm
    min} = 1.74$, $D_{\rm link} = 0.175\,\rm Mpc$, and $n_{\rm min} =
  5$.

Neither selection method shows an obvious bias to cluster
  ellipticity. However, the recovered shape of the clusters differs
  for both methods: VT tends to pick up more background galaxies at
  the edges of the clusters; by contrast, the galaxy members recovered
  by FOF are more centrally concentrated.

Cross-checking the output of the Voronoi Tessellations and the
  Friends-Of-Friends method eliminates spurious sources in the
  simulated cluster searches. However, low-level clustering within the
  background has not been taken into account.

The simulations yield completeness estimates as
  a function of redshift and cluster mass; these can be found in
  Fig.~\ref{completeness}. Furthermore, they provide us with a method
  of determining cluster mass, deduced from the number of galaxies
  found with the Friends-Of-Friends method and shown in Fig.~\ref{nz}.

\section*{Acknowledgments} 
The authors acknowledge STFC for financial support. Furthermore, we
thank Steve Rawlings for useful discussions, Dave Bonfield for his
efforts concerning the photometric redshift method, and all our
co-authors on Van Breukelen et al. (2006). Finally, we are grateful to
the referee for their helpful comments and suggestions to improve this paper.
\bibliographystyle{mn2e}

\section*{References}
\bib Abell G.\,O. 1958, ApJS, 3, 211
\bib Abell G.\,O., Corwin H.\,G. \& Olowin R.\,P.,1989, ApJS, 70, 1
\bib Bartelmann M., 1996, A\&A, 313, 697 
\bib Bolzonella M., Miralles J.-M., Pell{\'o} R., 2000, A\&A, 363, 476
\bib Botzler C.\,S., Snigula J., Bender R., Hopp U., 2004, MNRAS, 349, 425
\bib Bruzual G., Charlot S., 2003, MNRAS, 344, 1000
\bib Cole S. et al., 2001, MNRAS, 326, 255
\bib Delaunay B., 1934, Izvestia Akademii Nauk SSSR, Otdelenie
  Matematicheskikh i Estestvennykh Nauk, 7, 793
\bib Dolag K., Bartelmann M., Perrotta F., Baccigalupi C., Moscardini L., Meneghetti M., Tormen G., 2004, A\&A, 416, 853
\bib Ebeling H. \& Wiedenmann G., 1993, PhRvE, 47, 704
\bib  Gardner J.\,P., Sharples R.\,M., Carrasco B.\,E., Frenk C.\,S., 1996, MNRAS, 282L, 1
\bib Gladders M.\,D. \& Yee H.\,K.\,C., 2000, AJ, 120, 2148
\bib Gladders M.\,D. \& Yee H.\,K.\,C., 2005, ApJS, 157, 1
\bib Goto et al., 2002, AJ, 123, 1807
\bib Icke V., van de Weygaert R., 1987, A\&A, 184, 16
\bib Kiang T., 1966, Zeitschrift f{\" u}r Astrophysik, 64, 433
\bib Kim R.\,S.\,J. et al., 2002, AJ, 123, 20
\bib Koester B.\,P. et al., 2007(a), ApJ, 660, 221
\bib Koester B.\,P. et al., 2007(b), ApJ, 660, 239
\bib Lee D.\,T., Schachter B.\,J., 1980, Int. J. of Computer and Information Sci., 9, 219
\bib Lin Y.-T., Mohr J.\,J., Stanford S.\,A., 2004, ApJ, 610, 745
\bib Lobo C., Iovino A., Lazzati D., Chincarini G., 2000, A\&A, 360, 896
\bib Lopes P.\,A.\,A., deCarvalho R.\,R., Gal R.\,R., Djorgovski S.\,G., Odewahn S.\,C., Mahabal A.\,A., Brunner R.\,J., 2004, AJ, 128, 1017 
\bib Lucey J.\,R., 1983, MNRAS, 204, 33
\bib Miller C.\,J., Nichol R.\,C., G\'omez P.\,L., Hopkins A.\,H., Bernardi M., 2005, AJ, 130, 968
\bib Navarro J., Frenk C., White S., 1997, ApJ, 490, 493
\bib Postman M., Lubin L.\,M., Gunn J.\,E., Oke J.\,B., Hoessel J.\,G., Schneider D.\,P., Christensen J.\,A., 1996, AJ, 111, 615
\bib Ramella M., Boschin W., Fadda D., Nonino M., 2001, A\&A, 368, 776
\bib Ramella M., Geller M.\,J., Pisani A., da Costa L.\,N, 2002, AJ,123, 2976
\bib Rines K., Geller M.\,J., Kurtz M.\,J., Diaferio A., Jarrett T.\,H., Huchra J.\,P., 2001, ApJ, 561, L41
\bib Sutherland W., 1988, MNRAS, 234, 159
\bib Tucker D.\,L. et al., 2000, ApJS, 130, 237
\bib Van Breukelen C. et al., 2006, MNRAS, 373L, 26
\bib Van Breukelen C. et al., 2009, MNRAS, submitted
\bib Van Haarlem M.\,P., Frenk C.\,S., White S.\,D.\,M., 1997, MNRAS, 287, 817
\end{document}